\documentclass{elsart}
\usepackage{epsfig}
\def\mySpecialText{Belle DRAFT etapks_plb_final.tex, Ver 5.0}

 \def\myspecial#1{}                   %%  to print official version
\myspecial{!userdict begin /bop-hook{gsave 40 150 translate 90
rotate
    /Times-BoldItalic findfont 18 scalefont setfont
    0 0 moveto 0.70 setgray
    (\mySpecialText)
    show grestore}def end}

\def\DESepsf(#1 width #2){\epsfxsize=#2 \epsfbox{#1}}

\begin{document}

\pagestyle{empty}                                      %%%To be commented

\begin{frontmatter}

\begin{flushleft}\includegraphics[width=3.5cm]{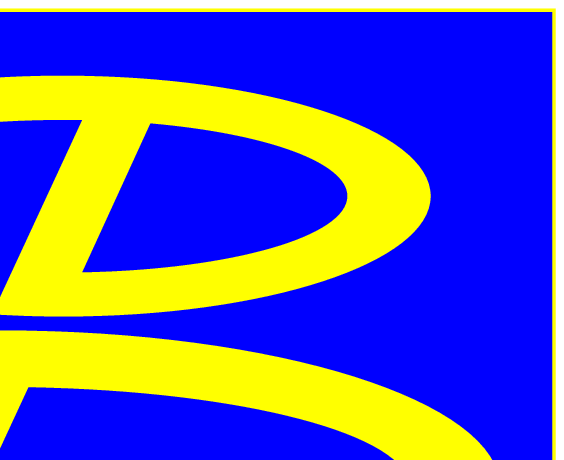}\end{flushleft}
\vspace{-3cm}
\vbox{\normalsize%
\noindent%
\rightline{\hfill {KEK Preprint 2002-66}}

\rightline{\hfill {Belle Preprint 2002-22}}

\rightline{\hfill {NTU-HEP-02-08}} }

\vspace{2.0cm}

%%%%%%%%%%%%%%%%%%%%%%%%%%%%%%%%%%%%%%%%%%%%%%%%%%
%  Title
%%%%%%%%%%%%%%%%%%%%%%%%%%%%%%%%%%%%%%%%%%%%%%%%%%

\title{\boldmath
Measurement of $CP$-Violating Parameters in $B\to \eta^\prime K$
Decays}
\date{\today}
\vspace{-1pc}

% Paper:    B -> eta' K
% Journal:  Physics Letters B
% Contacts: K.-F. Chen (kfjack@hep1.phys.ntu.edu.tw)
%           P. Yeh (pyeh@phys.ntu.edu.tw)
%           Y.B. Hsiung (hsiung@hep1.phys.ntu.edu.tw)
% Deadline: Thu Jul 10 2002 01:00 am JST
% Non-responding authors or those who said NO are commented out.
% ====================================================================
% Click the RELOAD button on your web browser to see the updated file.
% ====================================================================
% Use \input{author} to insert this material into your latex file.
\collab{Belle Collaboration}
  \author[Taiwan]{K.-F.~Chen}, % Taiwan
  \author[Osaka]{K.~Hara}, % Osaka
  \author[KEK]{K.~Abe}, % KEK
  \author[TohokuGakuin]{K.~Abe}, % TohokuGakuin
% \author[TIT]{N.~Abe}, % TIT
% \author[Niigata]{R.~Abe}, % Niigata
  \author[Tohoku]{T.~Abe}, % Tohoku
  \author[KEK]{I.~Adachi}, % KEK
  \author[Korea]{Byoung~Sup~Ahn}, % Korea
  \author[Tokyo]{H.~Aihara}, % Tokyo
  \author[Nagoya]{M.~Akatsu}, % Nagoya
% \author[Hiroshima]{M.~Asai}, % Hiroshima
  \author[Tsukuba]{Y.~Asano}, % Tsukuba
% ...
% Tsukuba

% ...
\address[KEK]{High Energy Accelerator Research Organization (KEK), Tsukuba, Japan}
% ...
\address[Korea]{Korea University, Seoul, South Korea}
% ...
\address[Nagoya]{Nagoya University, Nagoya, Japan}
% ...
\address[Taiwan]{National Taiwan University, Taipei, Taiwan}
% ...
\address[Osaka]{Osaka University, Osaka, Japan}
% ...
\address[TohokuGakuin]{Tohoku Gakuin University, Tagajo, Japan}
\address[Tohoku]{Tohoku University, Sendai, Japan}
\address[Tokyo]{University of Tokyo, Tokyo, Japan}
% ...
\address[Tsukuba]{University of Tsukuba, Tsukuba, Japan}
% ...
\thanks[Fermilab]{on leave from National Fermi Accelerator Laboratory, Batavia, IL, USA}
\thanks[NovaGorica]{on leave from Nova Gorica Polytechnic, Nova Gorica, Slovenia}
\thanks[Toronto]{on leave from University of Toronto, Toronto, ON, Canada}

%%%%%%%%%%%%%%%%%%%%%%%%%%%%%%%%%%%%%%%%%%%%%%%%%%
%
%  Abstract
%
%%%%%%%%%%%%%%%%%%%%%%%%%%%%%%%%%%%%%%%%%%%%%%%%%%
\begin{abstract}
We present measurements of $CP$-violating parameters in
$B^0(\overline{B}{}^0) \to \eta^\prime K_S^0$  and $B^{\pm} \to
\eta' K^{\pm}$ decays based on a 41.8 fb$^{-1}$ data sample
collected at the $\Upsilon(4S)$ resonance with the Belle detector
at the KEKB asymmetric-energy $e^{+} e^{-}$ collider. We fully
reconstruct one neutral $B$ meson as a $B^0(\overline{B}{}^0) \to
\eta^\prime K_S^0$ $CP$ eigenstate and identify the flavor of the
accompanying $B$ from its decay products. From the distribution of
proper time intervals between pairs of $B$ meson decay points, we
obtain the $CP$-violating asymmetry parameters ${\mathcal
S}_{\eta^\prime K_S^0} = 0.28\pm0.55(stat)^{+0.07}_{-0.08}(syst)$,
and ${\mathcal A}_{\eta^\prime K_S^0} =
0.13\pm0.32(stat)^{+0.09}_{-0.06}(syst)$. We also reconstruct
charged $B^{\pm} \to \eta^\prime K^{\pm}$ decays and determine a
direct-$CP$ violating asymmetry value of ${\mathcal
A}_{\eta^\prime K^\pm}=(-1.5\pm7.0(stat)\pm0.9(syst))\%$.

%\vskip1pc
%PACS: 13.25.Hw, 11.30.Er, 12.15.Hh
\end{abstract}


%{\renewcommand{\thefootnote}{\fnsymbol{footnote}}}

\end{frontmatter}
%@twocolumnfalse\endcsname
%
%\normalsize
%\vskip2pc ]                                                    % <---

%\newpage
\pagestyle{plain}

%%%%%%%%%%%%%%%%%%%%%%%%%%%%%%%%%%%%%%%%%%%%%%%%%%
%
%  Introduction
%
%%%%%%%%%%%%%%%%%%%%%%%%%%%%%%%%%%%%%%%%%%%%%%%%%%

In the Kobayashi and Maskawa (KM) model, $CP$ violation is
incorporated as an irreducible complex phase in the
weak-interaction quark mixing matrix~\cite{km}. Measurements of
sin$2\phi_1$, where $\phi_1 \equiv \pi -
$arg$(\frac{-V_{tb}^*V_{td}}{-V_{cb}^*V_{cd}})$, from $CP$
violation in $B^0 \to c\overline{c} K^0$ decays by the
Belle~\cite{Belle_sin2phi1} and BaBar~\cite{Babar_sin2phi1}
collaborations established $CP$ violation in the neutral $B$ meson
system that is consistent with KM expectations. Measurements of
sin$2\phi_1$ based on other decay modes, especially charmless
modes that are mediated by diagrams that contain virtual particle
loops, provide important tests of the KM model. In this letter we
describe the first measurement of $CP$-violating asymmetries in
the penguin-mediated decay $B^0 \to \eta^\prime K_S^0$, and an
improved measurement of the direct $CP$-violating asymmetry in the
decay $B^{\pm} \to \eta^\prime K^{\pm}$~\cite{ch-conj}.

The KM model predicts $CP$-violating asymmetries in the
time-dependent rates for $B^0$ and $\overline{B}{}^0$ decay to a
common $CP$ eigenstate, $f_{CP}$.  When an $\Upsilon(4S)$ decays
into a $B^0\overline{B}{}^0$ meson pair, the two mesons remain in
a coherent $p$-wave state until one of them decays. The decay of
one of the $B$ mesons at time $t_{\rm tag}$ to a final state,
$f_{\rm tag}$, which distinguishes between $B^0$ and
$\overline{B}{}^0$, projects the accompanying $B$ meson onto the
opposite $b$-flavor which decays to $\eta^\prime K_S^0$ at time
$t_{CP}$.  The decay rate has a time dependence given by
\begin{equation}
{\mathcal P}^q_{\eta^\prime K_S^0}(\Delta t) = {e^{-|\Delta
t|/\tau_{B^0}} \over 4\tau_{B^0}} \{ 1 + q\cdot A_{CP}(\Delta t)
\},
\label{eq:pdf_A_CP}
\end{equation}
\vspace{-1pc}
where $A_{CP}(\Delta t)$ is the time-dependent $CP$
asymmetry, $\tau_{B^0}$ is the $B^0$ lifetime, $\Delta
t~=~t_{CP}~-~t_{\rm tag}$, and the $b$-flavor $q=+1 (-1)$ when the
accompanying $B$ meson is a $B^0$($\overline{B}{}^0$).

Within the framework of the Standard Model (SM), the charmless
decay $B^0\to\eta^\prime K_S^0$ proceeds primarily via $b\to s$
penguin diagrams; there is a small contribution from a
color-suppressed $b\to u$ tree diagram, but that amplitude is
expected to be only a few percent of that for the $b\to s$
penguin~\cite{London&Soni,Ali_Kramer&Lu,Kou&Sanda}. Thus, $CP$
violation in this decay mode, to a good approximation, measures
$\phi_1$, and one can compare the result to the value measured for
$B^0 \to (c\overline{c}) K_S^0$. Phases from new physics in the
penguin loop could show up as a difference between the two
measured values~\cite{London&Soni,Moroi}. Since the branching
fraction of $B \to\eta^\prime K$ appears to be anomalously large
~\cite{Cleo_etapk_br,Babar_etapk_br,Belle_etapk_br}, this mode is
especially interesting to search for effects of an additional
phase besides $\phi_1$ due to physics beyond the
SM~\cite{London&Soni}.

The time-dependent $CP$ asymmetry can be expressed as
\vspace{-0.5pc}
\begin{eqnarray}
~~~~~~~~~ A_{CP}(\Delta t) & = & \frac{\Gamma(\overline{B}{}^0\to
\eta^\prime K_S^0) - \Gamma(B^0\to \eta^\prime
K_S^0)}{\Gamma(\overline{B}{}^0\to
\eta^\prime K_S^0) +\Gamma(B^0\to\eta^\prime K_S^0)} \nonumber       \\
~~~~~~~~~  & = & {\mathcal A}_{\eta^\prime K_S^0}\cos(\Delta m
\Delta t) + {\mathcal S}_{\eta^\prime K_S^0}\sin(\Delta m \Delta
t), \label{eq:A_CP}
\end{eqnarray}
where the $CP$-violating parameters ${\mathcal A}_{\eta^\prime
K_S^0}$ and ${\mathcal S}_{\eta^\prime K_S^0}$ are given by
\begin{equation}
   {\mathcal A}_{\eta^\prime K_S^0} = \frac{|\lambda|^2-1}{|\lambda|^2+1},  \;\;\;
   {\mathcal S}_{\eta^\prime K_S^0} = \frac{2Im\lambda}{1+|\lambda|^2},
\label{eq:par_asym}
\end{equation}
\vspace{-1pc}
in which $\lambda$ is a complex parameter that
depends on both $B^0$-$\overline{B}{}^0$ mixing and the decay
amplitude for $B^0(\overline{B}{}^0) \to\eta^\prime K_S^0$.  The
SM value for $|\lambda|$ is very nearly equal to the absolute
value of the ratio of the $\overline{B}{}^0$ to $B^0$ decay
amplitudes. Therefore $|\lambda| \not= 1$, or equivalently
${\mathcal A}_{\eta^\prime K_S^0} \not= 0$, indicates direct $CP$
violation. The direct $CP$ asymmetry, ${\mathcal A}_{\eta^\prime
K^\pm}$, in charged $B^\pm \to \eta^\prime K^\pm$ decays can also
be investigated from the time-integrated decay rates of $B^-$
versus $B^+$. To a good approximation the $CP$-violating
parameter, ${\mathcal S}_{\eta^\prime K_S^0}$, in the decay $B^0
\to \eta^\prime K_S^0$ is equal to the parameter $\sin2\phi_1$,
which can be directly compared with the value measured from $B^0
\to (c\overline{c})K^0$ decays.

%%%%%%%%%%%%%%%%%%%%%%%%%%%%%%%%%%%%%%%%%%%%%%%%%%
%
%  Experiment & Detector
%
%%%%%%%%%%%%%%%%%%%%%%%%%%%%%%%%%%%%%%%%%%%%%%%%%%

The measurement reported here is based on a 41.8 fb$^{-1}$ data
sample, which contains 44.8 million $B\overline{B}$ pairs,
collected with the Belle detector at the KEKB asymmetric energy
(3.5 GeV on 8 GeV) $e^+e^-$ collider~\cite{kekb} operating at the
$\Upsilon(4S)$ resonance. At KEKB, the $\Upsilon(4S)$ is produced
with a Lorentz boost of $\beta\gamma = 0.425$ nearly along the
electron beam direction ($z$). Since the $B^0$ and
$\overline{B}{}^0$ mesons are approximately at rest in the
$\Upsilon(4S)$ center-of-mass system, $\Delta t$ can be determined
from the displacement in $z$ between the $f_{CP}$ and
$f_{\rm{tag}}$ decay vertices: $\Delta t \simeq (z_{CP} -
z_{\rm{tag}})/\beta \gamma c \equiv \Delta z/\beta \gamma c$.

The Belle detector is a large solid-angle general purpose magnetic
spectrometer that consists of a three-concentric-layer silicon
vertex detector (SVD), a 50-layer central drift chamber (CDC), an
array of aerogel threshold \v Cerenkov counters (ACC),
time-of-flight scintillation counters (TOF), and an
electromagnetic calorimeter comprised of 8736 CsI(Tl) crystals
(ECL) located inside a superconducting solenoid coil that
provides a 1.5 T magnetic field. An iron flux-return located
outside of the coil is instrumented to detect muons and $K_L$ mesons
(KLM). The detector is described in detail
elsewhere~\cite{Belle_NIM}.

%%%%%%%%%%%%%%%%%%%%%%%%%%%%%%%%%%%%%%%%%%%%%%%%%%
%
%  $B \to \eta^\prime K$  Reconstruction
%
%%%%%%%%%%%%%%%%%%%%%%%%%%%%%%%%%%%%%%%%%%%%%%%%%%

The $B \to \eta^\prime K$ event selection closely follows the
method described in detail in a previously published report that
describes the branching ratio measurement~\cite{Belle_etapk_br};
the time-dependent $CP$ analysis is similar to that used for $B^0
\to \pi^+\pi^-$ and presented in Ref.~\cite{Belle_pipi}.

For the analysis reported here, the event selection is slightly
modified for the time-dependent measurement from what described in
Ref.~\cite{Belle_etapk_br} in order to retain more signal events
and to keep better control of systematic errors. Candidate $K_S^0
\to \pi^+\pi^-$ decays are reconstructed from pairs of oppositely
charged tracks that are constrained to a common vertex and have an
invariant mass that is within $\pm16$ MeV/$c^2$ of the nominal
$K_S^0$ mass. Two decay channels are used for $\eta^\prime$
reconstruction: $\eta^\prime \to \eta \pi^+\pi^-$
($\eta^\prime_{\eta\pi\pi}$) with $\eta \to \gamma \gamma$
($\eta_{\gamma\gamma}$); and $\eta^\prime \to \rho^0 \gamma$
($\eta^\prime_{\rho\gamma}$) with $\rho^0 \to \pi^+\pi^-$. To
increase the event yield, we omit the minimum transverse momentum
$p_t$ requirement for charged tracks that was used in
Ref.~\cite{Belle_etapk_br}. Instead, we require that all of the
tracks have associated SVD hits and radial impact parameters $|dr|
< 0.1$~cm projected on the $r$-$\phi$ plane. Better tracking
requirement improves the vertex determination and gives less bias
due to detector asymmetry. Particle identification information
from the ACC, TOF and CDC $dE/dx$ measurements are used to form a
likelihood ratio in order to distinguish pions from kaons with at
least 2.5$\sigma$ separation for laboratory momenta up to 3.5
GeV/$c$. Candidate photons from
$\eta_{\gamma\gamma}~(\eta^\prime_{\rho\gamma})$ decays are
required to be isolated and have $E_\gamma > 50~(100)$~MeV from
the ECL measurement. The invariant mass of $\eta_{\gamma \gamma}$
candidates is required to be between 500~MeV/$c^2$ and
570~MeV/$c^2$. A kinematic fit with an $\eta$ mass constraint is
performed using the fitted vertex of the $\pi^+\pi^-$ tracks from
the $\eta^\prime$ as the decay point. For
$\eta^\prime_{\rho\gamma}$ decays, the candidate $\rho^0$ mesons
are reconstructed from pairs of vertex-constrained $\pi^+\pi^-$
tracks with an invariant mass between 550 and 920~MeV/$c^2$. The
$\eta^\prime$ candidates are required to have a reconstructed mass
from 940 to 970~MeV/$c^2$ for the $\eta^\prime_{\eta\pi\pi}$ mode
and 935 to 975~MeV/$c^2$ for $\eta^\prime_{\rho\gamma}$ mode.
Charged $K^{\pm}$ candidates are selected for the decay $B^{\pm}
\to \eta^\prime K^{\pm}$ based on the particle identification
information described in Ref.~\cite{Belle_etapk_br}.

Candidate $B$ mesons are identified by combining $\eta^\prime$ and
$K_S^0$ ($K^\pm$) candidates to form the beam-constrained mass
$M_{\rm bc} \equiv \sqrt{(E^{\rm cms}_{\rm beam})^2 -(p^{\rm
cms}_B)^2}$, and the energy difference $\Delta E \equiv E^{\rm
cms}_B - E^{\rm cms}_{\rm beam}$, where $E^{\rm cms}_{\rm beam}$
is the center of mass (cms) beam energy (nominally $5.29$~GeV),
and $E^{\rm cms}_B$ and $p^{\rm cms}_B$ are the cms energy and
momentum of the $B$ candidate. The $M_{\rm bc}$ signal region is
defined as $M_{\rm bc}>5.27$~GeV/$c^2$; the $\Delta E$ signal
region depends on the mode: it is  $-0.1$~GeV $< \Delta E <$
$0.08$~GeV for $\eta^\prime_{\eta\pi\pi}$ and $|\Delta E| <$
$0.06$~GeV for $\eta^\prime_{\rho\gamma}$.

We extract the signal yields with a simultaneous unbinned
maximum-likelihood (ML) fit for both $\Delta E$ and $M_{\rm bc}$.
The signal distribution
% $F_{\rm sig}^{\Delta E,M_{\rm bc}}$,
is a product of a Gaussian function in $M_{\rm bc}$ and a Gaussian
plus a bifurcated Gaussian tail function in $\Delta E$. The means
and widths, as well as an overall normalization are the fitting
parameters. The shapes of the background distributions, described
below, are determined from sideband events in the $5.2$~GeV/$c^2$
$< M_{\rm bc} < 5.265$~GeV/$c^2$ and $|\Delta E| < 0.25$~GeV
region, but with $\Delta E <-0.12$~GeV or $\Delta E > 0.1$~GeV for
the $\eta^\prime_{\eta\pi\pi}$ mode and $|\Delta E|
> 0.08$~GeV for the $\eta^\prime_{\rho\gamma}$ mode;
the difference is due to slight differences in resolution in the
data as well as in the Monte Carlo simulation determined from the
higher statistics $\eta^\prime K^{\pm}$ mode.

%%%%%%%%%%%%%%%%%%%%%%%%%%%%%%
%
% Figures
%
%%%%%%%%%%%%%%%%%%%%%%%%%%%%%%
\begin{figure}[tb]
\begin{center}
\vspace{-1cm}
%\resizebox*{12cm}{12cm}{\includegraphics{eps/etapks-projections.eps}}
\includegraphics[width=12cm]{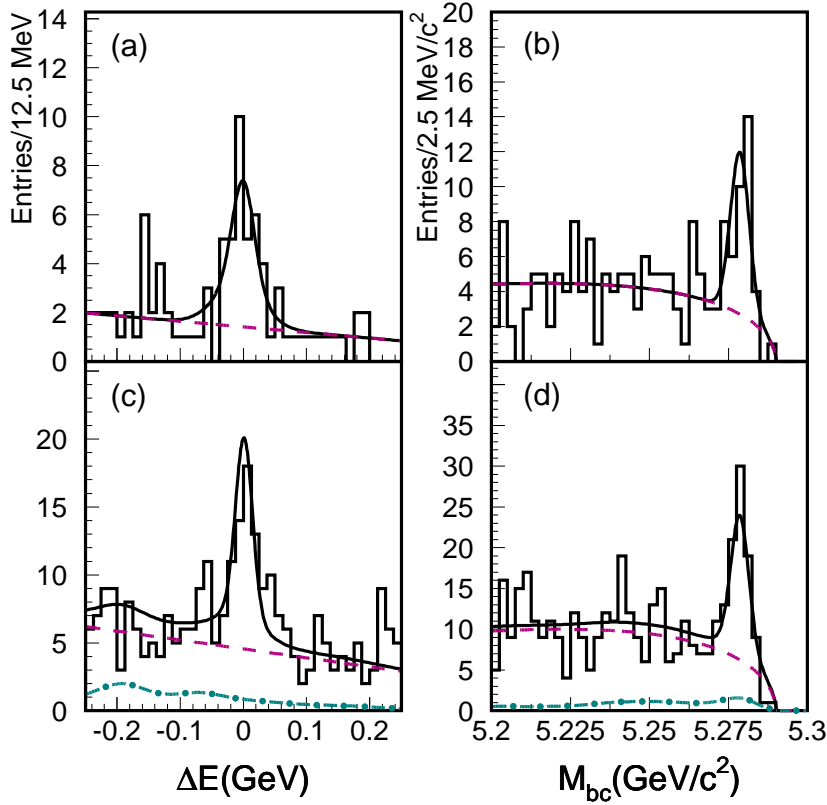}
\end{center}
\caption{The $\Delta E$ and $M_{\rm bc}$ distributions for the
$\eta^\prime K_S^0$ candidates (histogram) and the ML fit results
(solid curve); (a) and (b) show the $\eta^\prime\to\eta\pi\pi$
mode, (c) and (d) the $\eta^\prime\to\rho\gamma$ mode. The dashed
curves are the results of the fit for $q\overline{q}$ continuum
background; the dash-dot curves in (c) and (d) show the expected
$B\overline{B}$ background in the $\rho\gamma$ mode from the Monte
Carlo simulation.} \label{fig:demb-projections-ks}
\end{figure}

%%%%%%%%%%%%%%%%%%%%%%%%%%%%%%%%%%%%%%%%%%%%%%%%%%
%
%  Continuum  and BBbar Background
%
%%%%%%%%%%%%%%%%%%%%%%%%%%%%%%%%%%%%%%%%%%%%%%%%%%
The dominant backgrounds are from $e^+e^- \to q\overline{q}$
continuum events ($q = u,d,s,c$).  In this case, signal and
background events can be partially separated by the event
topology, which tends to be jet-like for $q\overline{q}$ continuum
events and nearly isotropic for $B\overline{B}$ events. We use
$|\cos\theta_{T}|$, the cosine of the angle between the thrust
axis of the $B$ candidate and that of the other particles in cms.
The requirement $|\cos\theta_{T}|<0.9$ rejects $~50\%$ of the
background while retaining $90\%$ of the signal events. This is
sufficient for the $\eta^\prime_{\eta\pi\pi}$ mode, which is
relatively clean.

For the $\eta^\prime_{\rho\gamma}$ channel, an additional cut is
applied using event shape variables. These variables include
$S_\perp$, which is the scalar sum of the transverse momenta of
all particles outside a $45^\circ$ cone around the candidate
$\eta^\prime$ direction divided by the scalar sum of their total
momenta, and five modified Fox-Wolfram moments~\cite{SFW}, all
combined into a single Fisher discriminant. In addition, we use
$\cos\theta_{B}$, the cosine of the angle between the $B$
candidate flight direction and the beam axis ($z$) in the cms, and
a $\rho$ helicity variable $H$, which is the cosine of the angle
between the $\pi^+$ momentum direction in the $\rho$ rest frame
and the $\rho$ momentum direction in the $\eta^\prime$ rest frame.
All of these variables are combined to form a likelihood ratio $LR
= L_S/(L_S + L_{q\overline{q}})$, where $L_{S(q\overline{q})}$ is
the product of signal ($q\overline{q}$) probability density
functions.  We determine $L_S$ from Monte Carlo (MC) and
$L_{q\overline{q}}$ from sideband data. We require $LR > 0.5$ for
the $\eta^\prime_{\rho\gamma}$ mode, which rejects $81\%$ of the
background and retains $80\%$ of the signal.

Sideband events are used to determine the shape of the continuum
background distributions for the $\eta^\prime_{\eta\pi\pi}$ mode
and $\eta^\prime_{\rho\gamma}$ modes separately. The $\Delta E$
shape is modelled by a linear background obtained from the $M_{\rm
bc}$ sideband. The $M_{\rm bc}$ shape is modelled by the empirical
function of Ref.~\cite{ARGUS_fn}.  From the ML fit we determine
the total background fraction in the signal region to be $40.4\%$
for the $\eta^\prime_{\eta\pi\pi} K_S^0$ mode and 56.5\% for the
$\eta^\prime_{\rho\gamma} K_S^0$ mode.  The backgrounds are
predominantly $q\overline{q}$ continuum events with small
contributions from other generic $B\overline{B}$ decays, mainly
due to charm daughter particles from $B$ or $\overline{B}$ decay.
These backgrounds, which are determined from a sample of MC
simulated generic $B$ meson decays, are negligible for the
$\eta^\prime_{\eta\pi\pi} K_S^0$ mode (smaller than $1\%$), but
contribute $14\%$ of the total background for the
$\eta^\prime_{\rho\gamma} K_S^0$ mode. The larger background for
the $\eta^\prime_{\rho\gamma} K_S^0$ mode is due to wide width of
$\rho$ mass and combinatorial background of $\gamma$ and $\rho$ to
form $\eta^\prime$ candidates.

%%%%%%%%%%%%%%%%%%%%%%%%%%%%%%%%%%%%%%%%%%%%%%%%%%
%
%  Yield and Signal efficiencies
%
%%%%%%%%%%%%%%%%%%%%%%%%%%%%%%%%%%%%%%%%%%%%%%%%%%

In the case of multiple candidates from the same event, we select
the candidate with the best $\chi^2$ value from the $\eta^\prime$
mass constrained fit. The signal efficiency is determined from MC
events to be $17.2~(16.5)\%$ for $\eta^\prime_{\eta\pi\pi} K_S^0$
($\eta^\prime_{\rho\gamma} K_S^0$). From the above-described ML
fit, we obtain $27.7^{+6.2}_{-5.5}$ signal events in the
$\eta^\prime_{\eta\pi\pi} K_S^0$ mode and $45.5^{+8.6}_{-7.9}$
events in the $\eta^\prime_{\rho\gamma}K_S^0$ mode; the results of
the fits are shown in Fig.~\ref{fig:demb-projections-ks}.
%The ratio of the signal yields in those two decay channels agrees
%with the ratio of the known branching fractions.

%%%%%%%%%%%%%%%%%%%%%%%%%%%%%%%%%%%%%%%%%%%%%%%%%%
%
%  Tag-side $B$ Reconstruction
%
%%%%%%%%%%%%%%%%%%%%%%%%%%%%%%%%%%%%%%%%%%%%%%%%%%

Leptons, $\Lambda$ baryons, and charged pions and kaons that are
not associated with the reconstructed $B^0 \to \eta^\prime K_S^0$
decay are used to identify the flavor of the accompanying $B$
meson. We apply the same method used for the Belle $\sin2\phi_1$
measurement~\cite{Belle_sin2phi1}. We use two parameters, $q$ and
$r$, to represent the tagging information: $q$ is a discrete
variable that corresponds to the sign of the $b$ quark charge and
has the value $q = +1$ for a $\overline{b}$ (i.e., $B^0$) tag, and
$q = -1$ for a $b$ ($\overline{B}{}^0$) tag;  $r$ is an
event-by-event, MC-determined flavor-tagging dilution factor that
ranges from $r=0$ (no flavor discrimination) to $r=1$ (unambiguous
flavor assignment). The value of $r$ is used to sort data into six
intervals of $r$, according to flavor purity, the corresponding
wrong tag fractions, $w_l (l = 1,6)$, and dilution factors
($1-2w_l$) are determined for each $r$ bin from data as described
in Ref.~\cite{Belle_sin2phi1}.
%
%directly from the data by the self-tagged semi-leptonic decay $B\to
%D^* l \nu$ and hadronic modes such as $B^0 \to D^{(*)-}\pi^+$ and
%$B^0\to D^{*-}\rho^+$ in the six $r$ regions specified as $0<r\leq
%0.25$, $0.25<r\leq 0.5$, $0.5<r\leq 0.625$, $0.625<r\leq 0.75$,
%$0.75<r\leq 0.875$, and $0.875<r\leq 1.0$. The b-flavor of the
%accompanying $B$ meson is assigned according to the flavor-tagging
%algorithm described above.

%%%%%%%%%%%%%%%%%%%%%%%%%%%%%%%%%%%%%%%%%%%%%%%%%%
%
%  Time Dependent Likelihood Fit and Result
%
%%%%%%%%%%%%%%%%%%%%%%%%%%%%%%%%%%%%%%%%%%%%%%%%%%

The vertex positions for the $\eta^\prime K_S^0$ and $f_{\rm tag}$
decays are reconstructed using tracks that have at least one
three-dimensional space point determined from associated
$r$-$\phi$ and $z$ hits in the same SVD layer and one or more
additional $z$ hits in the other layers. Each vertex position is
required to be consistent with the interaction point profile
smeared in the $r$-$\phi$ plane by the average transverse $B$
meson decay length. The $f_{\rm tag}$ vertex is determined from
all remaining well reconstructed tracks after the $B^0 \to
\eta^\prime K_S^0$ candidate tracks are removed. Tracks from other
$K_S^0$ candidates are not used. The MC simulation indicates that
the typical combined track-finding and vertex-finding efficiency
is 83$\%$ for $\eta^\prime K_S^0$ decays and 95$\%$ for the
$f_{\rm tag}$ decays. The typical vertex resolution (rms) is
147$~\mu$m (89$~\mu$m) for the $\eta^\prime_{\eta\pi\pi}$
($\eta^\prime_{\rho\gamma}$) mode on the $CP$ side, and 159$~\mu$m
for the $z_{\rm tag}$ on the tagging side.

The proper-time interval resolution for the signal, $R_{\rm
sig}(\Delta t)$, is obtained by convolving a sum of two Gaussians
(a {\it main} component due to the SVD vertex resolution and
charmed meson lifetimes, plus a {\it tail} component to account
for poorly reconstructed tracks) with a function that takes into
account the cms motion of the $B$ mesons. The fraction in the main
Gaussian is determined to be $0.97~\pm~0.02$ from a study of
control samples of fully reconstructed $B^0$
mesons~\cite{Belle_sin2phi1}. The means ($\mu_{\rm main}$,
$\mu_{\rm tail}$) and widths ($\sigma_{\rm main}$, $\sigma_{\rm
tail}$) of the Gaussians are calculated event-by-event from the
$\eta^\prime K_S^0$ and $f_{\rm tag}$ vertex fit error matrices
and the $\chi^2$ values from the fits~\cite{resolution_values}.
The background resolution $R_{q\overline{q}}(\Delta t)$ has the
same functional form but the parameters are obtained also
event-by-event from the $M_{\rm bc}$ and $\Delta E$ sideband
events. We obtain the lifetimes for neutral and charged $B$ mesons
for the $\eta^\prime K$ channels using the same procedure; the
results~\cite{lifetime_values} are consistent with the world
average values.

\begin{figure}[tpb]
\begin{center}
%\vspace{-5.5cm}
%\resizebox*{6cm}{15cm}{\includegraphics{eps/combine.eps}}
\includegraphics[height=18cm]{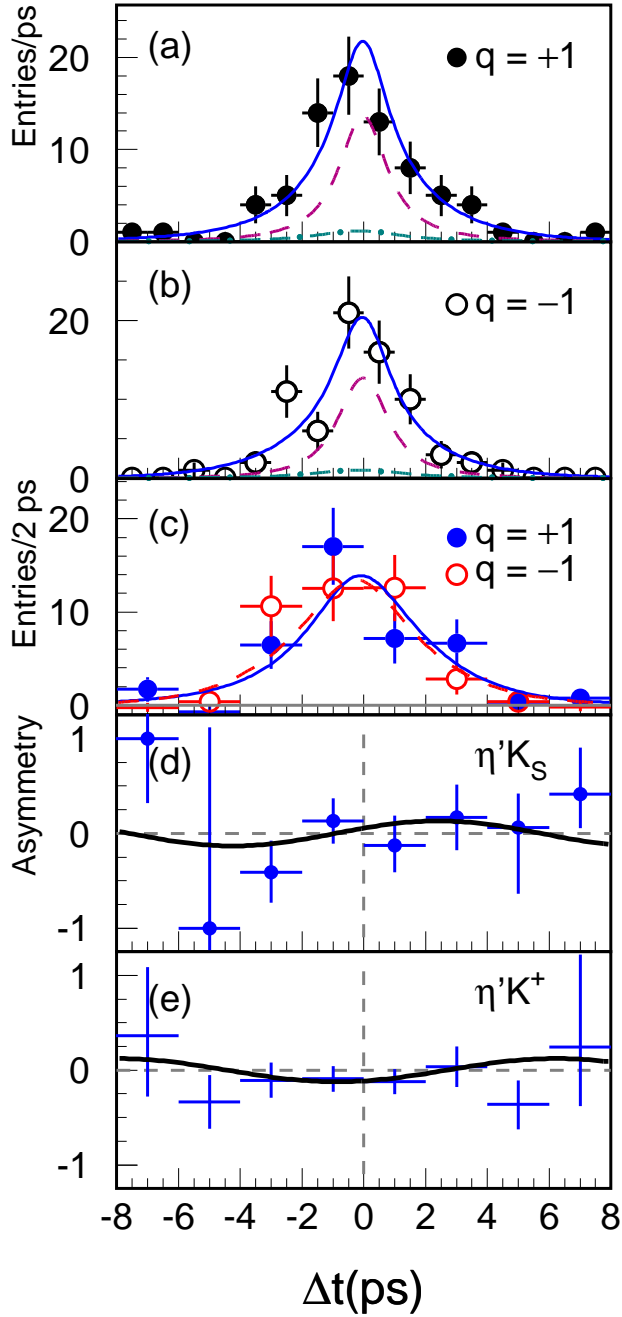}
\end{center}
\caption{The $\Delta t$ distributions for the $B^0 \to \eta^\prime
K_S^0$ candidates in the signal region: (a) candidates with
$q=+1$, i.e. the tag side is identified as $B^0$; (b) candidates
with $q=-1$; (c) $\eta^\prime K_S^0$ yields after background
subtraction; (d) the $CP$ asymmetry for $B^0 \to \eta^\prime
K_S^0$ after background subtraction; (e) the same asymmetry
calculated for $\eta^\prime K^{\pm}$ events. The curves in the
figures show the results of the unbinned maximum likelihood fit.
In (a) and (b), the solid, dashed and dash-dot curves are fit
results for the total, $q\overline{q}$ background and
$B\overline{B}$ background respectively. In (c) the solid (dashed)
curve is for the $q=+1$ ($q=-1$) signal fit.  In (d) and (e), the
solid curve is the fit result for ${\mathcal A}_{\eta^\prime
K_S^0}$ and ${\mathcal S}_{\eta^\prime K_S^0}$ while the dashed
curve corresponds to zero asymmetry. } \label{fig:combine_asym}
\end{figure}

After vertexing, we find 77 candidate events with $q=+1$ ($B^0$)
flavor tags and 74 candidate events with $q=-1$
($\overline{B}{}^0$) for $B^0 \to \eta^\prime K_S^0$.
Figures~\ref{fig:combine_asym}(a) and \ref{fig:combine_asym}(b)
show the observed $\Delta t$ distribution for the two samples.

%%%%%%%%%%%%%%%%%%%%%%%%%%%%%%
%
% Figures
%
%%%%%%%%%%%%%%%%%%%%%%%%%%%%%%

We determine the $CP$-violating parameters, ${\mathcal
A}_{\eta^\prime K_S^0}$ and ${\mathcal S}_{\eta^\prime K_S^0}$, by
performing an unbinned ML fit to the observed $\Delta t$
distributions. For perfect resolution, the probability density
function (pdf) for the signal, ${\mathcal P}^q_{\rm sig}$, as a
function of ${\mathcal A}_{\eta^\prime K_S^0}$ and ${\mathcal
S}_{\eta^\prime K_S^0}$ is given by Eqs.~\ref{eq:pdf_A_CP} and
\ref{eq:A_CP} with $q$ replaced by $q(1-2w_l)$ to take into
account the dilution due to mis-tagging. We fix $\tau_{B^0}$ and
$\Delta m_d$ at their world average values~\cite{PDG}. The pdf
used for the $q\overline{q}$ background distribution is
\begin{equation}
{\mathcal P}_{q\overline{q}}(\Delta t) = \frac{1}{2} \left\{
f_{\tau} \cdot \frac{e^{-|\Delta t|/\tau_{bg}}}{2\tau_{bg}} +
(1-f_{\tau})\cdot\delta(\Delta t) \right\},
\end{equation}
\vspace{-1pc} where $f_{\tau}$ is the background fraction with an
effective lifetime $\tau_{bg}$ and $\delta(\Delta t)$ is the Dirac
delta function.  We determine $f_{\tau}=0.064\pm0.022$ and
$\tau_{bg}=2.24\pm0.37$~ps from the sideband data. The pdf used to
account for the small generic $B\overline{B}$ background for the
$\eta^\prime_{\rho\gamma}$ mode is ${\mathcal
P}_{B\overline{B}}(\Delta t) = \delta(\Delta t)/2$. This
background is mainly due to mis-reconstructed secondary (charm)
decays, such as $D^{(*)0} \rho^{\pm}$, $D^{(*)\mp} \rho^{\pm}$,
$D^{(*)0} \pi^{\pm}$ and $D^{(*)} l \nu$, etc.~with $D$ mesons
decaying into a kaon and multiple pions or a $\rho$ meson.

We define the likelihood value for each event as:
\vspace{-0.5pc}
\begin{eqnarray}
P_i &=& \int \left\{ f^l_{\rm sig}{\mathcal P}^q_{\rm sig}(\Delta
t^\prime, w_l) R_{\rm sig}(\Delta t_i - \Delta t^\prime)
    + f^l_{q\overline{q}}{\mathcal P}_{q\overline{q}}(\Delta t^\prime)
    R_{q\overline{q}}(\Delta t_i - \Delta t^\prime) \right. \nonumber \\
    &+& \left. f^l_{B\overline{B}}{\mathcal P}_{B\overline{B}}(\Delta t^\prime)
    R_{B\overline{B}}(\Delta t_i - \Delta t^\prime) \right\} d\Delta
    t^\prime.
\end{eqnarray}
Here $f^l_k$ ($k = {\rm sig}, q\overline{q}$ or $B\overline{B}$
and $l = 1,6$) are the weighted probability functions determined
on an event-by-event basis as a function of $\Delta E$ and $M_{\rm
bc}$, properly normalized by the average signal and background
fractions in the fitting region for each $r$ interval for the
signal $\eta^\prime K_S^0$, $q\overline{q}$ and $B\overline{B}$
events, defined as
\vspace{-0.5pc}
\begin{eqnarray}
f^l_k(\Delta E,M_{\rm bc})
 &=& { g^l_k F_k^{\Delta E, M_{\rm bc}} \over
 {  g^l_{\rm sig} F^{\Delta E,M_{\rm bc}}_{\rm sig}
 + g^l_{q\overline{q}} F^{\Delta E,M_{\rm bc}}_{q\overline{q}}
 + g^l_{B\overline{B}} F^{\Delta E,M_{\rm bc}}_{B\overline{B}}  } }.
\end{eqnarray}
Here $g^l_{\rm sig} + g^l_{q\overline{q}} + g^l_{B\overline{B}} =
1$ and $ F_k^{\Delta E, M_{\rm bc}}$ are the shape functions for
the signal ($k=sig$), continuum background ($k=q\overline{q}$) and
$B\overline{B}$ background ($k=B\overline{B}$).
 The average event fractions, $g^l_k$, are measured for
$\eta\pi\pi$ and $\rho\gamma$ separately in each $r$ interval.

In the fit, ${\mathcal A}_{\eta^\prime K_S^0}$ and ${\mathcal
S}_{\eta^\prime K_S^0}$ are free parameters that are determined by
maximizing the likelihood function ${\mathcal L}=\prod_i P_i$,
where the product is over all $B^0 \to \eta^\prime K_S^0$
candidates. After maximizing the combined likelihood, the $CP$
asymmetry parameters are determined from a total of 72.9
$\eta^\prime K_S^0$ signal events (37.3 $B^0$- and 35.6
$\overline{B}{}^0$-tags) to be:
%after subtracting the background:
\begin{center}
${\mathcal S}_{\eta^\prime K_S^0} = 0.28\pm0.55(stat)^{+0.07}_{-0.08}(syst)$, \\
${\mathcal A}_{\eta^\prime K_S^0} =
0.13{\pm0.32}(stat)^{+0.09}_{-0.06}(syst)$.
\end{center}
%
%and
%\begin{center}
%$\sin2\phi_1 = 0.30\pm0.54(stat)\pm0.07(syst)$.
%\end{center}
%

Figures~\ref{fig:combine_asym}(a) and (b) show the $\Delta t$
distribution for $B^0$- and $\overline{B}{}^0$-tagged events
together with the fit curves; the background-subtracted $\Delta t$
distributions for signal-only are shown in
Fig.~\ref{fig:combine_asym}(c). The errors on data points in
Fig.~\ref{fig:combine_asym}(c) are statistical only and do not
include the error associated with the subtracted background
obtained by the fit. The background-subtracted time-dependent $CP$
asymmetry between the $B^0$- and $\overline{B}{}^0$-tagged events
is shown as a function of $\Delta t$ in
Fig.~\ref{fig:combine_asym}(d), with the result of the fit for
${\mathcal S}_{\eta^\prime K_S^0}$ and ${\mathcal A}_{\eta^\prime
K_S^0}$ superimposed.

%%%%%%%%%%%%%%%%%%%%%%%%%%%%%%%%%%%%%%%%%%%%%%%%%%
%
%  Systematics
%
%%%%%%%%%%%%%%%%%%%%%%%%%%%%%%%%%%%%%%%%%%%%%%%%%%

The systematic errors are summarized in Table~\ref{tab:syserror}
for ${\mathcal S}_{\eta^\prime K_S^0}$ and ${\mathcal
A}_{\eta^\prime K_S^0}$.  The dominant sources for ${\mathcal
S}_{\eta^\prime K_S^0}$ are due to uncertainties in the signal and
background determination ($\Delta E$ and $M_{\rm bc}$ pdf's, event
fractions and background shape), resolution functions, wrong tag
fractions, vertexing, and the physics parameters ($\tau_{B^0}$ and
$\Delta m_d$). For ${\mathcal A}_{\eta^\prime K_S^0}$, the
uncertainties in the signal and background determination and the
vertexing are the leading components. We determine the systematic
error due to physics parameters by repeating the fit after varying
those parameters within their error taken from the world
average~\cite{PDG}. The systematic errors for wrong tag fractions,
resolution functions, signal and background pdf's are estimated by
repeating the fit after varying the parameters by $\pm 1\sigma$
determined from the data or MC. The systematic error due to
vertexing is studied by varying the cut on vertex $\chi^2$ or
removing it and then repeat the fit.

A number of checks are also performed.  We analyze the
$\eta^\prime K^\pm$ sample in the same way as $\eta^\prime K_S^0$.
The raw asymmetry for the $B^\pm \to \eta^\prime K^\pm$ candidates
is shown in Fig.~\ref{fig:combine_asym}(e).  A fit to 230
candidate events yields ${\mathcal S} = 0.11 \pm 0.29$ and
${\mathcal A} = -0.27 \pm 0.17$, consistent with no asymmetry, as
expected. We also examine the event yields and the $\Delta t$
distributions for $B^0$- and $\overline{B}{}^0$-tagged events in
the $\Delta E$ and $M_{\rm bc}$ sideband region.  We find no
significant difference between the two samples in both the
$\eta^\prime K^\pm$ and $\eta^\prime K_S^0$ modes. The average raw
asymmetry from the sideband data is $-0.009 \pm 0.014$, which is
consistent with zero and indicates no bias.

%%%%%%%%%%%%%%%%%%%%%%%%%%%%%%%%%%%%%%%%%%%%%%%%%%
%
%  eta' K^+
%
%%%%%%%%%%%%%%%%%%%%%%%%%%%%%%%%%%%%%%%%%%%%%%%%%%
%
\begin{figure}[tpb]
\begin{center}
%\vspace{-2cm}
%\resizebox*{12cm}{6cm}{\includegraphics{eps/de-acp.eps}}
\includegraphics[width=12cm]{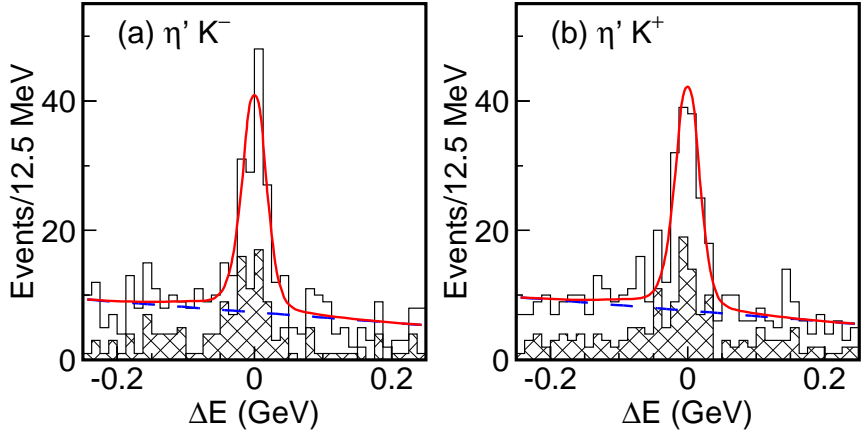}
\end{center}
\caption{The $\Delta E$ projection plots for the $\eta^\prime K^\pm$
candidates: (a) for $\eta^\prime K^-$, (b) for $\eta^\prime K^+$.  The
cross hatched histograms indicate the $\eta\pi\pi$ channel; the solid
histograms are the sum of both $\eta^\prime$ decay channels
($\eta\pi\pi$ and $\rho\gamma$). The curves are the fitted backgrounds
(dashed) and the sum of signal and background (solid). }
\label{fig:de-acp-kx}
\end{figure}

We also analyze the self-tagged $\eta^\prime K^{\pm}$ event sample
to search for a direct $CP$ asymmetry. Here we do not use
flavor-tag information from the other $B$ meson. The $\eta^\prime
K^{\pm}$ sample is divided into $\eta^\prime K^+$ and $\eta^\prime
K^-$ samples. Since the $CP$ asymmetry (${\mathcal A}_{\eta^\prime
K^\pm}=(N(B^-)-N(B^+))/(N(B^-)+N(B^+)$) for $\eta^\prime K^{\pm}$
is time independent, we use the same event selection and analysis
procedure as described in Ref.~\cite{Belle_etapk_br}. A
simultaneous ML fit to $M_{\rm bc}$ and $\Delta E$ is performed
for each sub-sample. The fitted numbers of signal events in the
$\eta^\prime_{\eta\pi\pi}$ ($\eta^\prime_{\rho\gamma}$) mode are
$66.0^{+10.0}_{-9.2}$ ($73.7^{+14.3}_{-13.5}$) for $B^-$ decays
and $66.6^{+10.1}_{-9.3}$ ($77.5^{+10.8}_{-10.0}$) for $B^+$
decays, respectively.  The number of produced $B^-$ and $B^+$
events are obtained by maximizing the product of the likelihoods
for each submode, as shown in Fig.~\ref{fig:de-acp-kx}. From the
fit we obtain ${\mathcal A}_{\eta^\prime K^\pm} = (-0.4\pm11.0)\%$
for the $\eta^\prime_{\eta\pi\pi}$ mode and $(-2.5\pm 10.0)\%$ for
$\eta^\prime_{\rho\gamma}$; here the errors are statistical only.
Since the systematic errors on $\eta^\prime$ reconstruction and
the number of $B\overline{B}$ events cancel in the ratio, the
systematic uncertainty in ${\mathcal A}_{\eta^\prime K^\pm}$ comes
mainly from the reconstruction efficiency of charged kaons and the
ML fit. The asymmetry in the $K^{\pm}$ efficiency is studied using
inclusive charged kaons in the same kinematic range as the signal.
The uncertainty due to fitting is measured by varying the
parameters of the fit functions. We find the systematic errors in
${\mathcal A}_{\eta^\prime K^\pm}$ are $0.85\%$ from $K^\pm$
reconstruction and $0.47\%$ from the ML fit.   The combined
${\mathcal A}_{\eta^\prime K^\pm}$ for the $B^\pm \to \eta^\prime
K^\pm$ decay is determined to be
\begin{center}
${\mathcal A}_{\eta^\prime K^\pm} =
(-1.5\pm7.0(stat)\pm0.9(syst))\%$,
\end{center}
which is consistent with zero. Combining the errors in quadrature
and assuming a Gaussian distribution, we find the $90\%$
confidence level interval is $-0.13 < {\mathcal A}_{\eta^\prime
K^\pm} < 0.10$, which is a factor of two more restrictive than our
previous measurement~\cite{Belle_etapk_br}.

%%%%%%%%%%%%%%%%%%%%%%%%%%%%%%%%%%%%%%%%%%%%%%%%%%
%
%  Summary
%
%%%%%%%%%%%%%%%%%%%%%%%%%%%%%%%%%%%%%%%%%%%%%%%%%%

In summary, we have measured the $CP$ asymmetry parameters in
$B^0(\overline{B}{}^0) \to \eta^\prime K_S^0$ and $B^{\pm} \to
\eta' K^{\pm}$ decays based on a 41.8 fb$^{-1}$ data sample
collected with the Belle detector. The result for
decay-time-integrated direct $CP$ asymmetry, ${\mathcal
A}_{\eta^\prime K^\pm}= (-1.5\pm7.0(stat)\pm0.9(syst))\%$, is
small and consistent with zero.  Our results for the
time-dependent asymmetry parameters ${\mathcal S}_{\eta^\prime
K_S^0} = 0.28\pm0.55(stat)^{+0.07}_{-0.08}(syst)$ and ${\mathcal
A}_{\eta^\prime K_S^0} =
0.13{\pm0.32}(stat)^{+0.09}_{-0.06}(syst)$ are the first
measurements of $CP$ asymmetry parameters related to $\phi_1$ with
a charmless $B^0$ decay mode. In the SM, to a good approximation
the value of ${\mathcal S}_{\eta^\prime K_S^0}$ should be equal to
$\sin2\phi_1$ measured in $B^0 \to c\overline{c}K^0$ decays, where
the current world average is $0.78\pm0.08$~\cite{sin2phi1_avg}.
With a much larger data sample, we will significantly reduce the
uncertainty in ${\mathcal S}_{\eta^\prime K_S^0}$ and impose tight
constraints on phases from new physics beyond the Standard Model.

%%%%%%%%%%%%%%%%%%%%%%%%%%%%%%%%%%%%%%%%%%%%%%%%%%
%
%  Acknowledgement
%
%%%%%%%%%%%%%%%%%%%%%%%%%%%%%%%%%%%%%%%%%%%%%%%%%%

We wish to thank the KEKB accelerator group for the excellent
operation of the KEKB accelerator. We acknowledge support from the
Ministry of Education, Culture, Sports, Science, and Technology of
Japan and the Japan Society for the Promotion of Science; the
Australian Research Council and the Australian Department of
Industry, Science and Resources; the National Science Foundation
of China under contract No.~10175071; the Department of Science
and Technology of India; the BK21 program of the Ministry of
Education of Korea and the CHEP SRC program of the Korea Science
and Engineering Foundation; the Polish State Committee for
Scientific Research under contract No.~2P03B 17017; the Ministry
of Science and Technology of the Russian Federation; the Ministry
of Education, Science and Sport of the Republic of Slovenia; the
National Science Council and the Ministry of Education of Taiwan;
and the U.S.\ Department of Energy.

%%%%%%%%%%%%%%%%%%%%%%%%%%%%%%%%%%%%%%%%%%%%%%%%%%
%
%  Reference
%
%%%%%%%%%%%%%%%%%%%%%%%%%%%%%%%%%%%%%%%%%%%%%%%%%%

%\newpage
\vspace{1cm}
%%%%%%%%%%%%%%%%%%%%%%%%%%%%%%%%%%%%%%%%%%%%%%%%%%
%
%  Tables
%
%%%%%%%%%%%%%%%%%%%%%%%%%%%%%%%%%%%%%%%%%%%%%%%%%%
\begin{table}[h]
  \caption{\label{tab:syserror} Summary of systematic errors.}
  %\begin{ruledtabular}
    \begin{tabular}{lcc}
      \hline
      Source   & $\delta({\mathcal S}_{\eta^\prime K_S^0})$   &  $\delta({\mathcal A}_{\eta^\prime K_S^0})$ \\
      \hline
      $\tau_{B^0}$ and $\Delta m_d$ & ${}^{+0.029}_{-0.034}$    &  ${}^{+0.006}_{-0.003}$  \\
      Wrong tag fractions      & ${}^{-0.028}_{-0.025}$         &  ${}^{+0.013}_{-0.012}$  \\
      Resolution functions     & ${}^{+0.037}_{-0.033}$         & ${}^{+0.011}_{-0.008}$   \\
      Vertexing                & ${}^{+0.018}_{-0.033}$         & ${}^{+0.062}_{-0.031}$   \\
      Event fractions          & ${}^{+0.029}_{-0.024}$         & ${}^{+0.012}_{-0.010}$   \\
      Background shapes        & ${}^{+0.013}_{-0.017}$         & ${}^{+0.004}_{-0.003}$   \\
      $\Delta E$ and $M_{\rm bc}$ pdf's & ${}^{+0.032}_{-0.028}$  & ${}^{+0.063}_{-0.054}$  \\
      \hline
      Total                    & ${}^{+0.07}_{-0.08}$           & ${}^{+0.09}_{-0.06}$ \\
      \hline
    \end{tabular}
  %\end{ruledtabular}
\end{table}


\begin{thebibliography}{99}
\bibitem{ch-conj}{
    Charge conjugate modes are implicitly included throughout the
    paper unless explicitly stated.
}
\bibitem{resolution_values}{
    Typical values are $\mu_{\rm main}=-0.24$~ps, $\mu_{\rm tail}=-0.37$~ps
    and $\sigma_{\rm main}=1.06$~ps, $\sigma_{\rm tail}=3.22$~ps.
}
\bibitem{lifetime_values}{We obtain
    $\tau_{B^\pm}=1.54\pm0.14(stat)$~ps for the
    $\eta^\prime K^\pm$ mode (the world average~\cite{PDG} is $1.653\pm0.028$~ps) and
    $\tau_{B^0}=1.58\pm0.30(stat)$~ps for the
    $\eta^\prime K_S^0$ mode (the world average is $1.548\pm0.032$~ps).
}
\end{thebibliography}
\end{document}